\documentclass[12pt,onecolumn]{IEEEtran}
\usepackage{unicode}
\usepackage{graphicx}
\usepackage{algorithmic}
\usepackage{algorithm} 
\usepackage{theorem}
\usepackage{subfigure}
\usepackage[top=1.5in, left=0.75in, right=0.75in, bottom=0.75in]{geometry}
\theoremheaderfont{\upshape\bfseries}
\theorembodyfont{\itshape}
\theoremstyle{plain}

\newtheorem{Rem}{Remark}

\newtheorem{Def}{Definition}
\newtheorem{Theorem}{Theorem}

\begin{document}


\abovecaptionskip5pt \belowcaptionskip5pt

\title{Stability and Distributed Power Control in MANETs with Outages and Retransmissions}
\author{Anastasios Giovanidis, \textit{Member, IEEE} and S\l awomir Sta\'nczak, \textit{Member, IEEE}}
\maketitle
\IEEEpeerreviewmaketitle
 
\footnotetext{This research is supported by the FET Open Project - FP6 - IST-034413 NetReFound. 

Part of this work has been presented during the 5th Int. Workshop on Resource Allocation, Cooperation and Competition in Wireless Networks (RAWNET/WNC3),  Seoul, Korea,  June 2009 (part of WiOpt'09).

The authors are with the Fraunhofer German-Sino Mobile Communications Lab, Heinrich-Hertz-Institut, Einsteinufer 37, D-10587 Berlin, Germany. Tel/Fax: +493031002-860/-863, e-mail: \{giovanidis, stanczak\} @ hhi.de and with the Technical University of Berlin, Heinrich Hertz Chair for Mobile Communications, Werner-von-Siemens-Bau (HFT 6), Einsteinufer 25, D-10587 Berlin, Germany.}


\begin{abstract}
In the current work the effects of hop-by-hop packet loss and retransmissions via ARQ protocols are investigated within a Mobile Ad-hoc NET-work (MANET). Errors occur due to outages and a success probability function is related to each link, which can be controlled by power and rate allocation. We first derive the expression for the network's capacity region, where the success function plays a critical role. Properties of the latter as well as the related maximum goodput function are presented and proved. A Network Utility Maximization problem (NUM) with stability constraints is further formulated which decomposes into (a) the input rate control problem and (b) the scheduling problem. Under certain assumptions problem (b) is relaxed to a weighted sum maximization problem with number of summants equal to the number of nodes. This further allows the formulation of a non-cooperative game where each node decides independently over its transmitting power through a chosen link. Use of supermodular game theory suggests a price based algorithm that converges to a power allocation satisfying the necessary optimality conditions of (b). Implementation issues are considered so that minimum information exchange between interfering nodes is required. Simulations illustrate that the suggested algorithm brings near optimal results.
\end{abstract}


\newpage

\section{Introduction}
The current work considers a Mobile Ad-hoc NETwork (MANET) where data flows entering from a set of source nodes should be routed to their destinations. In such networks a major concern is the maximum set of incoming rates that can be supported, since interference is the bottleneck. If a utility function is related to each incoming flow a very interesting problem is to maximize the sum of all utilities under the constraint that the queues of all nodes remain stable. Such problems have been addressed in \cite{TassMultihop}, \cite{NeelyRouting}, \cite{NeelyHetero}, \cite{LinShroffCDC04}, \cite{StanczakBook2} and algorithms that optimally adapt the incoming rates and the transmission powers of each node have been suggested. 

In the current work we are interested in bringing these models a step further and investigate how the stability regions and the optimal policies for congestion control, routing and power allocation vary, when the queues of each node use ARQ protocols to repeat transmissions of erroneous packets due to outages. In the current literature, investigations already addressing the network utility maximization (NUM) problem with erroneous transmissions through the links consider mainly fixed routing. In \cite{PapMANET} the model does not consider queueing aspects and a NUM problem with rate-outage constraints per link is approximately solved. In \cite{LeeChiangReliable} the effect of end-to-end error probability is included in the utility of each source. The same problem with average power and reliability requirements is posed and algorithmically solved in \cite{NeillWNUMAllerton}. Furthermore, in \cite{LauraICC09} a single hop ad-hoc network with outages is considered where a solution for joint admission control, rate and power allocation is derived based on a stochastic game formulation. Other contributions that investigate the effect of retransmissions in MANETs incorporating Random Access MAC protocols include \cite{McCabeCDC08}, \cite{ElAzouziMultihopRout}.

Motivated by a comment in \cite{XiaoBoyd04} where it is stated that "\textit{in practical communication systems, the link capacity should be defined appropriately, taking packet loss and retransmissions into account, hence the flow conservation law holds for goodputs instead of rates}", and after a presentation of the model under study in Section II, we derive in Section III the goodput capacity region. The success probability for the transmission over a wireless link depends on the entire power allocation and the scheduled transmission rate. We constrain our investigations to functions with specific properties presented in section IV, where it is shown that these also hold for the expression with Rayleigh fading \cite{KaBo02}. The NUM problem naturally decomposes in Section V into the input rate control and the scheduling problem. 

At this point a major challenge is to achieve a decentralized solution of the second problem. This is always possible of course for the case of parallel channels (see also \cite{XiaoBoyd04} and \cite{ChiangINFOCOM04}). Algorithmic suggestions can be found in \cite{NeelyRouting} for zero-full power allocations and in \cite{ChenLowChiang06} by solving a maximum weighted matching problem over a conflict graph. In our work fully distributed implementation is achieved by approaching the second problem with the arsenal of supermodular game theory in section VI - an idea appearing in \cite{HuangBerryHonigGame} and \cite{PowerPriceSMG02} - and result to the suggestion of a price based algortithm in VI.D and VI.E that achieves almost optimal solutions with minimum information exchange between the nodes. Simulation results are finally presented in section VII.

\section{Model under Study}
\begin{figure}[h]
\centering
\includegraphics[width=9cm]{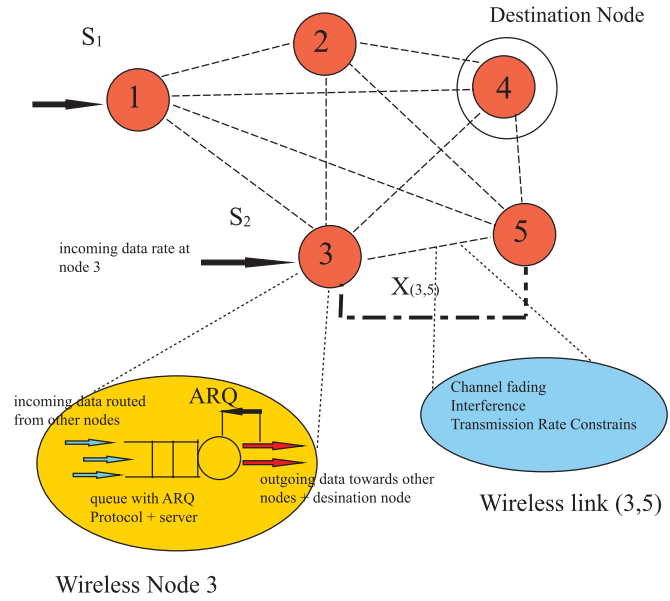}
\caption{An example of the wireless network with a single commodity $d = 4$. Detail of node 3}
\label{ITW09:fig1}
\end{figure}

We consider a wireless network consisting of $N$ nodes $\mathcal{N}=\left\{1,\ldots,N\right\}$, while $\mathcal{L}$ is the set of all possible $L = N\cdot \left(N-1\right)$ links. The time is divided into slots of equal duration $T$ (normalized to $1$) and $t=\left\{0,1,\ldots\right\}$ is the time index. Data flows enter the network at source nodes and are removed at destination nodes $\mathcal{D}=\left\{1,\ldots,D\right\}$. The set of data flows (\textit{commodity flows}) injected into the network at a source node with a predefined destination is denoted by $\mathcal{S}=\left\{1,\ldots,S\right\}$. The routes of the data flows through the network are not fixed. Furthermore, each link $l\in \mathcal{L}$ is characterized by an origin node $b\left(l\right)$ (begin) and an end node $e\left(l\right)$ (end). At each node $n$, a total of $D$ buffers - one for each commodity flow - are reserved (see also Fig.\ref{ITW09:fig1}).

In the general case investigated, each node $n\in\mathcal{N}$ chooses at slot $t$ a power $p_l\left(t\right)$ as well as a rate $\mu_l\left(t\right)$ to transmit data through link $l$, as long as $n=b\left(l\right)$. The total transmission rate of packets through link $l$ at some time slot $t$ is the sum of the transmission rates of the individual commodities sharing the link meaning $\mu_l\left(t\right)=\sum_{d=1}^D\mu_l^d\left(t\right)$. Scheduled packets of variable length $\mu_l^d\left(t\right)$ for each commodity $d$ are combined into a common packet of length $\mu_l\left(t\right)$ and sent through the link. The resulting long packet may be received at node $e\left(l\right)$ with errors due to fading and interference. The probability of successful transmission is then a real valued function $q\left(\vec{p},\mu\right):\mathbb{R}^{N\cdot \left(N-1\right)}_+ \times  \mathcal{M} \rightarrow \left[0,1\right]$ of the entire power allocation at slot $t$, $\vec{p}\left(t\right)\in \Pi\subseteq \mathbb{R}^{N\cdot \left(N-1\right)}_+$ and the scheduled rate $\mu_l\left(t\right)$. The set of all possible scheduling rates is denoted by $\mathcal{M}$. The nodes have power restrictions, e.g. $\forall n,\ \sum_{l:n=b\left(l\right)}p_l\left(t\right)\leq P_n$ and $\Pi$ is the convex compact set of all feasible power allocations.

Examples of such success probability functions for flat block-fading channels can be found using the outage probability definition \cite{GiovCISS08}. Given an $SINR_l$ threshold value $\gamma\left(\mu_l\right)= e^{\mu_l}-1$ of link $l$ (we often simply write $\gamma_l:= \gamma\left(\mu_l\right)$)

\begin{eqnarray}
\label{ITW09:eq01}
q_l\left(\vec{p},\mu_l\right) = \mathbb{P}\left(SINR_l\left(\vec{p}\right)\geq e^{\mu_l}-1\right), & & SINR_l\left(\vec{p}\right) = \frac{G_{ll}F_{ll}p_l}{\sum_{j\neq l}{G_{lj}F_{lj}p_j}+\sigma^2_{e\left(l\right)}}
\end{eqnarray}

%
where $Y_{lj}$ stands for $Y_{b\left(j\right)e\left(l\right)}$, $\mu_l$ is the scheduled transmission rate through the link, $G_{b\left(j\right)e\left(l\right)}$ is the slow varying path gain and $F_{b\left(j\right)e\left(l\right)}$ is the associated flat fading component of the channel. For the case of Rayleigh/Rayleigh fading (meaning Rayleigh slow fading for both the desired and interference signals), a closed form expression of (\ref{ITW09:eq01}) can be found in \cite{KaBo02} and \cite{PapOut}

\begin{eqnarray}
\label{ITW09:eq03}
q_l\left(\vec{p},\mu_l\right) = \exp\left(\frac{-\sigma^2 \gamma_l}{G_{ll}p_l}\right)\prod_{j\neq l}\left(1+\frac{\gamma_l G_{lj}p_j}{G_{ll}p_l}\right)^{-1}.
\end{eqnarray}
Observe that the success functions used imply that only the channel fast fading \textit{statistics} are known and the nodes have no other instantaneous channel state information (CSI) over the fading gains, except - possibly - the slow varying path gains. The actual amount of data transmitted through each link equals $\mu_l\left(t\right)\cdot X_l\left(t\right)$. $X_l\left(t\right)$ is a binary random variable which equals $1$ for success (with prob. $q_l$) and $0$ for failure (with prob. $1-q_l$). The expected transmission rate through link $l$ is then

\begin{eqnarray}
\label{ITW09:eq03-1}
g_l\left(\vec{p},\mu_l\right) := \mu_l\cdot q_l\left(\vec{p},\mu_l\right)
\end{eqnarray}
and is called the \textit{goodput} of link $l$ \cite{AhBa03}, \cite{BetteshIT06}. Furthermore, in the analysis that follows we often encounter a quantity called \textit{maximum goodput} defined as (see \cite{AhBa03} and \cite{GiovSCC08} for parallel Rayleigh fading channels)

\begin{eqnarray}
\label{ITW09:eq03-2}
g_l\left(\vec{p}\right) = \max_{\mu_l\in\mathcal{M}}\mu_l\cdot q_l\left(\vec{p},\mu_l\right).
\end{eqnarray}

In case a packet of length $\mu_l\left(t\right)$ is received at node $e\left(l\right)$ with errors, we assume that this can always be detected during decoding. When reception is correct an ACK is fed back otherwise a NAK signal is transmitted to $b\left(l\right)$ via a reliable zero-delay wireless feedback link. In the latter case the packets of all transmitted commodities are then not removed from the buffer but wait for a future retransmission (Stop-and-Wait ARQ) under some new scheduling decision. The queue evolution for each node $n$ and commodity flow $d$ at slot $t$, is given by

\begin{eqnarray}
\label{ITW09:eq04}
u_n^d\left(t+1\right) & = & \left[u_n^d\left(t\right)- \sum_{k:n=b\left(k\right)} \mu_k^d\left(t\right) X_k\left(t\right)\right]^+ + \sum_{l:n=e\left(l\right)} \mu_l^d\left(t\right) X_l\left(t\right) + \alpha_n^d\left(t+1\right).
\end{eqnarray}

The success probability of the transmission through link $l$ is equal for all commodities $d$, since it depends on the sum rate $\mu_l$. In the expression (\ref{ITW09:eq04}), $\sum_{k:n=b\left(k\right)} \mu_k^d\left(t\right) X_k\left(t\right)$ is the actual outgoing data ("actual" meaning "error free") from node $n$, $\sum_{l:n=e\left(l\right)} \mu_l^d\left(t\right) X_l\left(t\right)$ is the actual incoming data from links $l\in\mathcal{L}: n = e\left(l\right)$, $n\neq d$ and $\alpha_n^d\left(t+1\right)$ is the amount of commodity $d$ bits arriving exogenously to the network at node $n$ during $t$.

We associate each incoming flow to the network at node $n$ with destination $d\in \mathcal{D}$, $\alpha_n^d = \alpha_s$, with a utility function $U_s:\mathbb{R}_+\rightarrow\mathbb{R}_+$. The utility function takes as argument the average incoming data rate $\mathbb{E}\left(\alpha_s\right) = x_s$ and is non-decreasing, strictly concave and continuously differentiable over the range $x_s\geq 0$ (\textit{elastic traffic}, \cite{KellyElastic}). The utilities describe the satisfaction received by transmitting data from node $s\in\mathcal{S}$ to $d\in\mathcal{D}$.

The aim here is to find an incoming rate vector $\vec{x} = \left(x_1,\ldots,x_S\right)$ to maximize the sum of the utilities $\sum_{s\in\mathcal{S}}U_s\left(x_s\right)$ subject to the constraint that the system remains stable and furthermore explicitly provide the stabilizing scheduling policy $\forall \vec{x}\in\Lambda$. $\Lambda$ denotes the \textit{capacity region} of the system, the largest set of $\vec{x}$ for which the system remains stable. Formally we write 

\begin{eqnarray}
\label{ITW09:eq06}
\mathbf{max}\ \ \sum_{s\in \mathcal{S}}U_s\left(x_s\right) & \mathbf{subject\ to}\ \ \vec{x}\in\Lambda.
\end{eqnarray}

\section{Network Capacity Region and Variations with Dropping Packet Decisions}
The problem posed so far is similar to the models investigated in \cite{NeelyRouting}, \cite{NeelyHetero}, \cite{LinShroffCDC04} and \cite{ChenLowChiang06}. Due to the occurence of errors and the use of retransmissions, the capacity region of the model under investigation is definitely reduced and has a different expression compared to the works mentioned.

\begin{Theorem}
\label{ITW09:Th1}
The capacity region $\Lambda$ of the wireless network under study is the set of all non-negative vectors $\vec{x} = \left(x_1,\ldots,x_S\right)$ 
such that there exist \textit{multicommodity \textbf{goodput} flow variables} $\left\{g_l^d\right\}_{l\in\mathcal{L}}^{d\in\mathcal{D}}$, satisfying
\begin{itemize}
\item $g_l^d\geq 0,\ \forall l\in\mathcal{L},d\in\mathcal{D}$ and $g_l^d = 0$ if $e\left(l\right) = d$
\item $\forall n\in\mathcal{N},d\in\mathcal{D}$: $\sum_{l:e(l) = n}g_l^d + x_n^d\leq \sum_{k:b(k) = n}g_k^d$
\item $\sum_{d\in\mathcal{D}}g_l^d\leq g_l$, $\vec{g} = \left\{g_l\right\}\in\Gamma$, where $\Gamma = \mathbf{co}\left(\hat{\Gamma}\right)$ and
\end{itemize}

\begin{eqnarray}
\label{ITW09:eq07}
\hat{\Gamma} = \bigcup_{\vec{p}\in\Pi} \left\{ \vec{g}\in\mathbb{R}_+^{N\cdot{\left(N-1\right)}}: \forall l\in\mathcal{L},g_l\leq \bar{\mu}_l\left(\vec{p}\right)\cdot q_l\left(\vec{p},\bar{\mu}_l\left(\vec{p}\right)\right),\ \bar{\mu}_l\left(\vec{p}\right) = \arg\max_{\mu_l\in \mathcal{M}}\mu_l\cdot q_l\left(\vec{p},\mu_l\right)\right\}.
\end{eqnarray}
\end{Theorem}

\begin{proof}
Similar to the derivation of the network capacity region in \cite{NeelyRouting} and can be found in \cite{GiovRAWNET09}.
\end{proof}

In the above $\left\{g_l^d\right\}_{l\in\mathcal{L}}^{d\in\mathcal{D}}$ is the $D\cdot N\cdot(N-1)$ size vector of goodput flow variables for all commodities through the network. An optimal policy achieving stability for all vectors within $\Lambda$ is a variation of the well-known backpressure policy \cite{TassMultihop}, \cite{NeelyRouting} where goodputs replace the rate vectors. This is named here \textbf{goodput} backpressure policy. We further denote with $\Gamma$ the \textit{goodput region} of the network, which equals the convex hull ($\mathbf{co}$) of $\hat{\Gamma}$ given in (\ref{ITW09:eq07}). Comparing this region to the ones appearing in \cite{NeelyRouting} and \cite{ChiangINFOCOM04} the rate-power mapping $r_l\left(\vec{p}\right) = \log\left(1+SINR_l\left(\vec{p}\right)\right)$ is replaced here by the maximum goodput-power mapping $g_l\left(\vec{p}\right)$.

Let us now assume that the nodes can decide, in addition to the transmission power $p_l $ and rate $\mu_l$ over the link $l\in\mathcal{L}:n = b\left(l\right)$, whether the possibly erroneous packet at time slot $t$ should be dropped or should be held in the node's queues and wait to be retransmitted at the next time slot $t+1$. We use the binary decision variable $A_l\left(t\right)$ taking values $A_l\left(t\right) = 0$ for dropping decision and $A_l\left(t\right) = 1$ for a decision to continue. The single queue evolution will be the same as in (\ref{ITW09:eq04}) where $X_l$ (and similarly $X_k$) should be replaced by the expression $1-A_l\left(t\right)\left(1-X_l\left(t\right)\right)$ which equals $X_l\left(t\right)$ when $A_l\left(t\right) = 1$ and $1$ when $A_l\left(t\right) = 0$.

If the decisions on dropping are randomized, with a fixed probability of dropping per link equal to $1-\delta_l\in\left[0, 1\right]$ (and hence $\mathbb{E}A_l\left(t\right) = \delta_l$), the network capacity region $\Lambda_{\vec{\delta}}$, $\vec{\delta} = \left(\delta_1,\ldots,\delta_L\right)$, will be the same as in Theorem \ref{ITW09:Th1} with a modification on the region $\hat{\Gamma}$. In this case we have that 

\begin{eqnarray}
\label{ITW09:eq08}
\hat{\Gamma}_{\vec{\delta}} = \bigcup_{\vec{p}\in\Pi}\left\{ \vec{g}\in\mathbb{R}_+^{N\cdot{\left(N-1\right)}}: \forall l\in\mathcal{L}, g_l\leq \bar{\mu}_l^{\delta_l} \left(1-\delta_l\cdot \left(1-q_l\left(\vec{p},\bar{\mu}_l^{\delta_l}\right)\right)\right)\right\}
\end{eqnarray}
$\bar{\mu}_l^{\delta_l}:= \bar{\mu}_l\left(\vec{p},\delta_l\right) = \arg\max_{\mu_l\in \mathcal{M}}\mu_l \cdot \left(1-\delta_l\cdot \left(1-q_l\left(\vec{p},\mu_l\right)\right)\right)$. Choice of the vector $\vec{\delta} = \vec{1} := \left(1,\ldots,1\right)$ results in the region of Theorem \ref{ITW09:Th1} where no dropping takes place, while for $\vec{\delta} = \vec{0} := \left(0,\ldots,0\right)$, dropping always takes place after an erroneous transmission and this provides the maximum network capacity region with $\hat{\Gamma}_{\vec{\delta}}$ equal to 

\begin{eqnarray}
\label{ITW09:eq09}
\hat{\Gamma}_{\vec{\delta} = \vec{0}} = \left\{ \vec{g}\in\mathbb{R}_+^{N\cdot{\left(N-1\right)}}: \forall l\in\mathcal{L},g_l\leq \mu_l^* \right\}
\end{eqnarray}
where $\mu_l^* = \arg\max_{\mu_l\in\mathcal{M}} \mu_l$ is the maximum allowable transmission rate per link. We can then obtain different regions $\Lambda_{\vec{\delta}}$ between these two extremes by varying the dropping probabilities per link. To understand why this is important suppose that a network user transmitting a data flow with source node $s\in\mathcal{S}$ has a higher data rate than that offered by the actual error free network capacity region $\Lambda_{\vec{\delta}=\vec{1}}$. We may then vary the vector $\vec{\delta} $ so that the network will fit the requirements of the user. Of course the \textit{average} rate of \textit{correctly} transmitted packets through the network will not change. What will happen is that, instead of removing part of the user's packets at entering the network (admission control), the network will offer per link at least one chance for all packets to be correctly transmitted through the network, hence will be able to provide \textit{unreliable} service to the \textit{entire} required high data rate, with index of reliablity $\vec{\delta}$.

\section{Properties of the success function and the maximum goodput function}

The success probability function $q_l\left(\vec{p},\mu_l\right)$ for transmission over link $l\in\mathcal{L}$ considered in this work, has the following properties\footnote{The game-theoretic notation $q_l\left(p_l,\vec{p}_{-l},\mu_l\right)$ is often used, where $\vec{p}_{-l}$ is the entire power vector excluding the $l$-th element $p_l$.}.

\begin{itemize}
\item \textbf{P.1} $q_l$ is strictly \textit{increasing} in $p_l$ and the $\log$ of the function is concave in $p_l$

\item \textbf{P.2} $q_l$ is strictly \textit{decreasing} and \textit{convex} in $p_k,\ \forall k\neq l,k\in \mathcal{L}$

\item \textbf{P.3} $q_l$ is strictly \textit{decreasing} in $\mu_l$

\item \textbf{P.4} The $\log$ of the function has \textit{increasing differences} for the pair of variables $\left(p_l,\mu_l\right)$ meaning that

\begin{eqnarray}
\label{ITW09:eq10}
\log q_l\left(p_l^+,\vec{p}_{-l},\mu_l\right) - \log q_l\left(p_l,\vec{p}_{-l},\mu_l\right) & \leq & 
\log q_l\left(p_l^+,\vec{p}_{-l},\mu_l^+\right) - \log q_l\left(p_l,\vec{p}_{-l},\mu_l^+\right)
\end{eqnarray}
where $p_l^+\geq p_l$ and $\mu_l^+\geq \mu_l$.

\item \textbf{P.5} The $\log$ of the function has \textit{increasing differences} for each pair of variables $\left(p_l,p_j\right),\forall j\neq l$. 
The differences are \textit{constant} for all pairs $\left(p_i,p_j\right)$, where $i\neq j$ and $i,j\in\mathcal{L}\backslash\left\{l\right\}$.
\end{itemize}

The last property actually implies - using \cite[Corollary 2.6.1]{Topkis} - that the function is $\log$-\textit{supermodular}. By property \textbf{P.4} a positive change on the transmission power $p_l$ has a greater impact on the increase of the (logarithm of the) success probability, the higher the rate of transmission. If we e.g. transmit with 16-QAM modulation, an increase of power by $\Delta p_l>0 $ will increase $\log q$ much more than in the case of transmission with BPSK.

\begin{Theorem}
\label{ITW09:Th2}
The success probability function for the Rayleigh/Rayleigh fading case, given in (\ref{ITW09:eq03}) satisfies properties \textbf{P.1-P.5}. 
\end{Theorem}

\begin{proof}
For the proof, the expressions (\ref{ITW09:partial1}) - (\ref{ITW09:partial4}) of first and second order partial derivatives are required. Specifically, from (\ref{ITW09:partial1}) and (\ref{ITW09:partial2}) the function is increasing in $p_l$ and decreasing in $p_j$ (strictly if $p_l\geq P_l^{\min}>0$ and same for $j$). From (\ref{ITW09:partial3}) the logarithm of the function is concave in $p_l$. The convexity in \textbf{P.2} comes directly from the partial derivative of (\ref{ITW09:partial2}) over $p_j$ which is easily shown to be positive. \textbf{P.3} is shown in (\ref{ITW09:partial5}), whereas \textbf{P.4} comes directly by derivating (\ref{ITW09:partial5}) w.r.t. $p_l$. Finally, \textbf{P.5} is a direct consequence of the fact that - in (\ref{ITW09:partial4}) - $\frac{\partial^2 \log q_l\left(\vec{p},\mu_l\right)}{\partial p_l\partial p_j} \geq 0$ and $\frac{\partial^2 \log q_l\left(\vec{p},\mu_l\right)}{\partial p_i\partial p_j} = 0$ (see \cite[p.42]{Topkis}).
\end{proof}
\begin{eqnarray}
\label{ITW09:partial1}
\frac{\partial q_l\left(p_l,\vec{p}_{-l},\mu_l\right)}{\partial p_l} = &  q_l\left(p_l,\vec{p}_{-l},\mu_l\right)\cdot\left[ \frac{\sigma^2\gamma_l\left(\mu_l\right)}{G_{ll}p_l^2} +\sum_{j\neq l}\frac{1}{\frac{G_{ll}p_l^2}{\gamma_l\left(\mu_l\right)G_{lj}p_j}+p_l}\right] & \geq 0\\
\label{ITW09:partial2}
\frac{\partial q_l\left(p_l,\vec{p}_{-l},\mu_l\right)}{\partial p_j} = & -q_l\left(p_l,\vec{p}_{-l},\mu_l\right)\cdot \frac{1}{\frac{G_{ll}p_l}{\gamma_l\left(\mu_l\right)G_{lj}}+p_j}& \leq 0
\end{eqnarray}
\begin{eqnarray}
\label{ITW09:partial5}
\frac{\partial q_l\left(p_l,\vec{p}_{-l},\mu_l\right)}{\partial \mu_l} = &  q_l\left(p_l,\vec{p}_{-l},\mu_l\right)\cdot\left[ \frac{-\sigma^2 e^{\mu_l}}{G_{ll}p_l} -\sum_{j\neq l}\frac{e^{\mu_l}G_{lj}p_j}{G_{ll}p_l+\gamma_l\left(\mu_l\right)G_{lj}p_j}\right] & \leq 0\\
\label{ITW09:partial3}
\frac{\partial^2 \log q_l\left(p_l,\vec{p}_{-l},\mu_l\right)}{\partial p_l^2} = & -\frac{2\sigma^2\gamma_l\left(\mu_l\right)}{G_{ll}p_l^3} -\sum_{j\neq l}\frac{\frac{2p_lG_{ll}}{\gamma_l\left(\mu_l\right)G_{ljp_j}}+1}{\left(\frac{G_{ll}p_l^2}{\gamma_l\left(\mu_l\right)G_{lj}p_j}+p_l\right)^2}& \leq 0\\
\label{ITW09:partial4}
\frac{\partial^2 \log q_l\left(p_l,\vec{p}_{-l},\mu_l\right)}{\partial p_l\partial p_j} = & \frac{\frac{G_{ll}}{\gamma_l\left(\mu_l G_{lj}\right)}}{\left(\frac{G_{ll}p_l}{\gamma_l\left(\mu_l\right)G_{lj}}+p_j\right)^2}& \geq 0.
\end{eqnarray}

Using the above properties we can derive important properties for the maximum goodput function in (\ref{ITW09:eq03-2}), which as seen in (\ref{ITW09:eq07}) plays a critical role in the definition of the system capacity region.

\begin{Theorem}
\label{ITW09:Th3}
If the success probability function satisfies \textbf{P.1-P.5} then the \textit{maximum goodput} function in (\ref{ITW09:eq03-2}) has the following properties (where $\bar{\mu_l}\left(\vec{p}\right) = \arg\max_{\mu_l\in \mathcal{M}} \mu_l q_l\left(\vec{p},\mu_l\right)$)
\begin{itemize}
\item \textbf{P'.1} $g_l\left(\vec{p}\right)$ is \textit{strictly increasing} in $p_l$

\item \textbf{P'.2} $g_l\left(\vec{p}\right)$ is \textit{strictly decreasing} and \textit{convex} in $p_k,\ \forall k\neq l$

\item \textbf{P'.3} $\bar{\mu_l}\left(\vec{p}\right)$ is \textit{non-decreasing} in $p_l$

\item \textbf{P'.4} $\bar{\mu_l}\left(\vec{p}\right)$ is \textit{non-increasing} in $p_k,\ \forall k\neq l$

\end{itemize}

\end{Theorem}

\begin{proof}
Proofs of \textbf{P'.1-P'.4} are found in Appendix A.
\end{proof}

\begin{figure}[h]
\centering
\includegraphics[width=8cm]{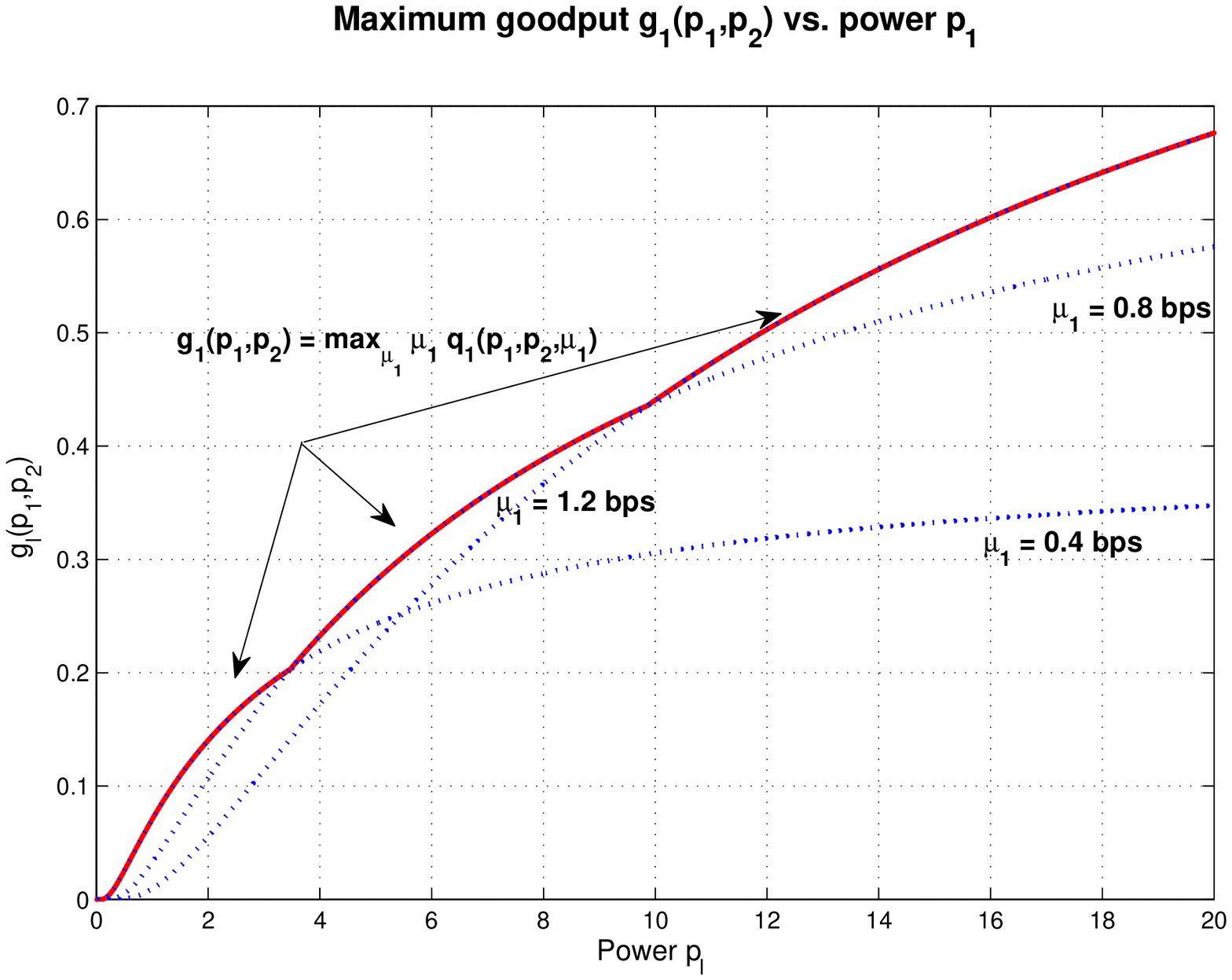}
\includegraphics[width=8cm]{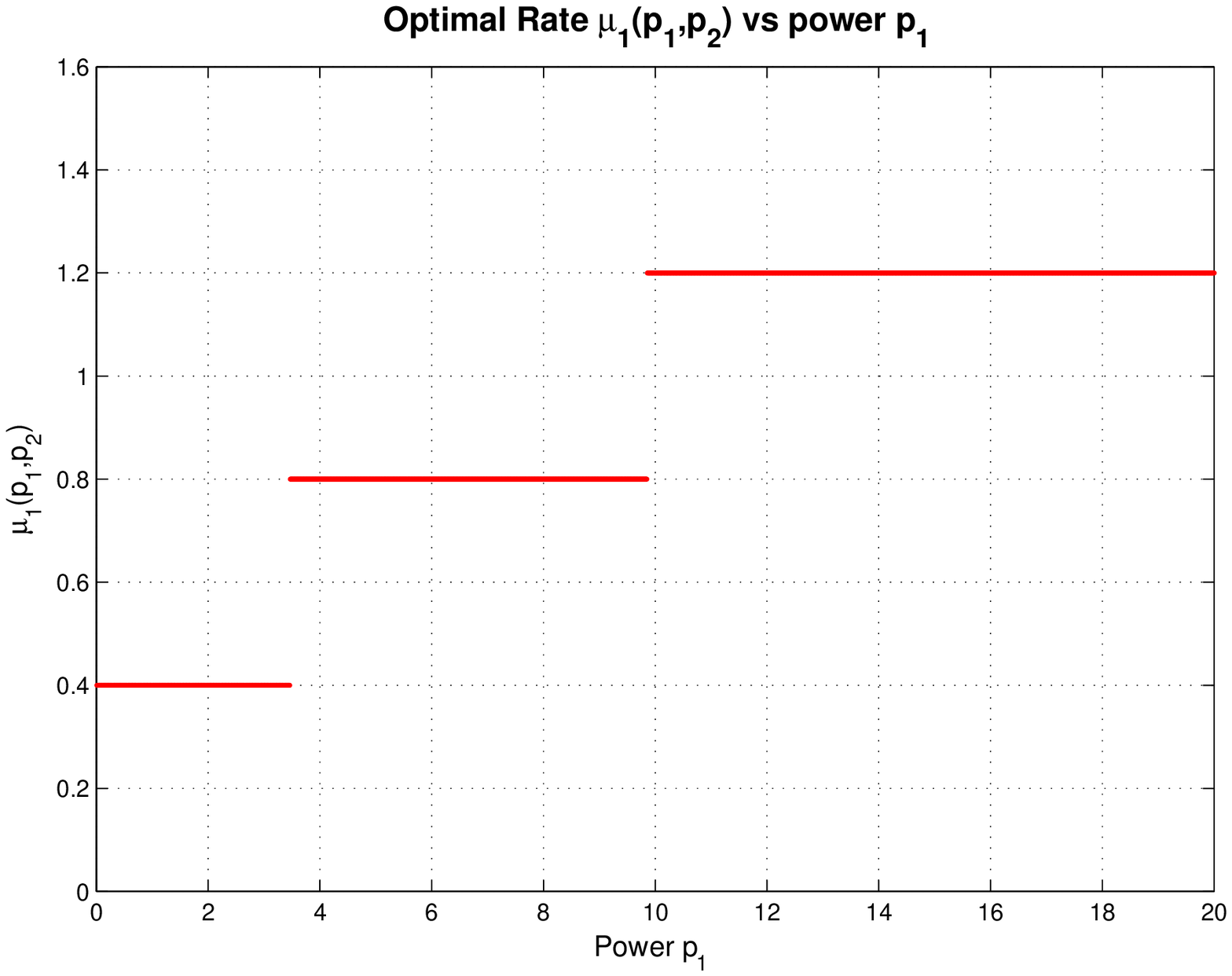}
\caption{We simulate a 2-user Rayleigh/Rayleigh fading channel with set of rates $\mathcal{M} = \left\{0.4,0.8,\ldots,2\right\}$ and $p_2 = 5$ Watt, $P_1 = 20$ Watt. Properties \textbf{P'.1} and \textbf{P'.3} are illustrated a. Maximum Goodput $g_1(p_1,p_2)$ vs power $p_1$, b. Optimal rate $\bar{\mu}_1\left(p_1,p_2\right)$ vs power $p_1$.}
\label{ITW09:fig2}
\end{figure}

\begin{figure}[h]
\centering
\includegraphics[width=8cm]{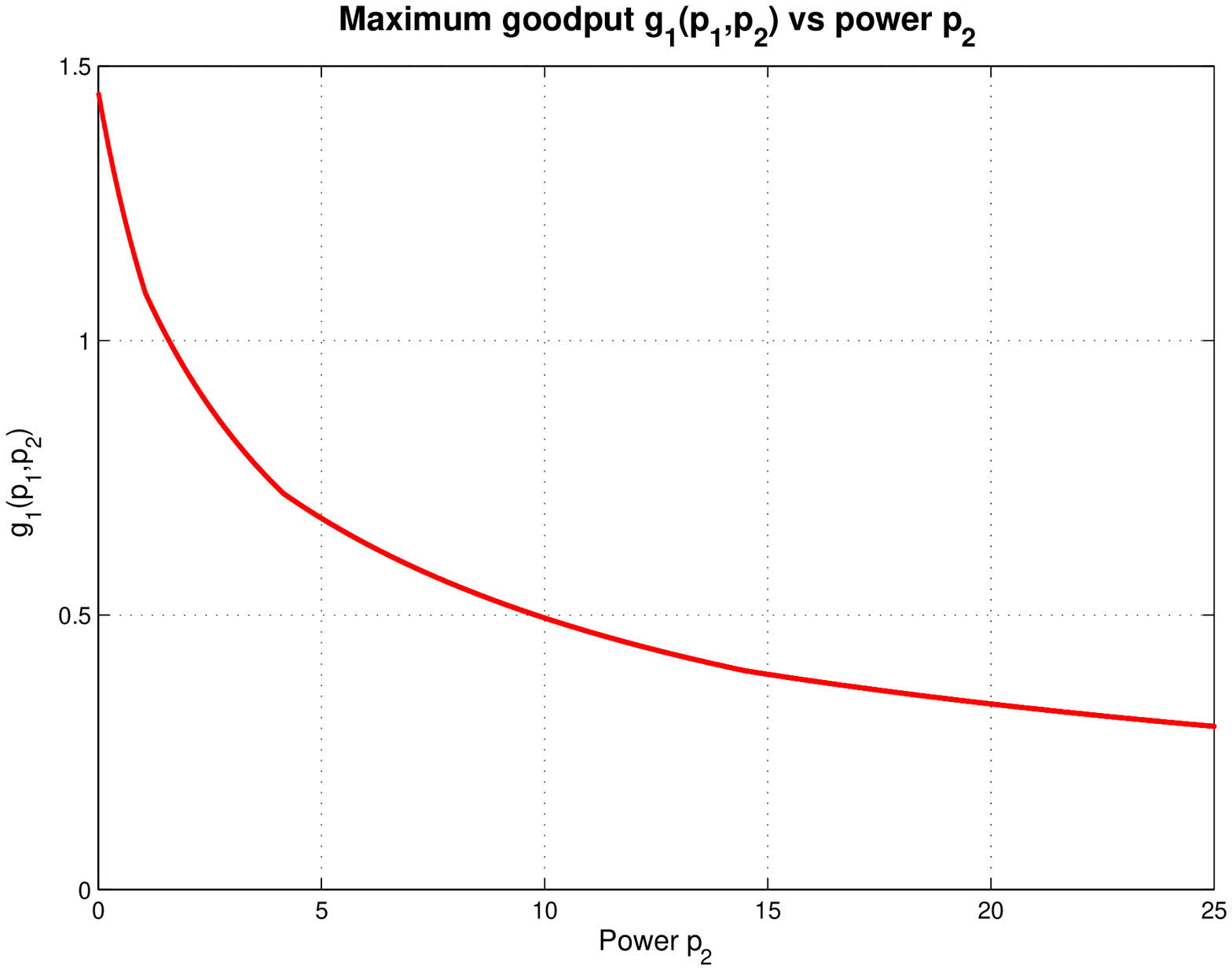}
\includegraphics[width=8cm]{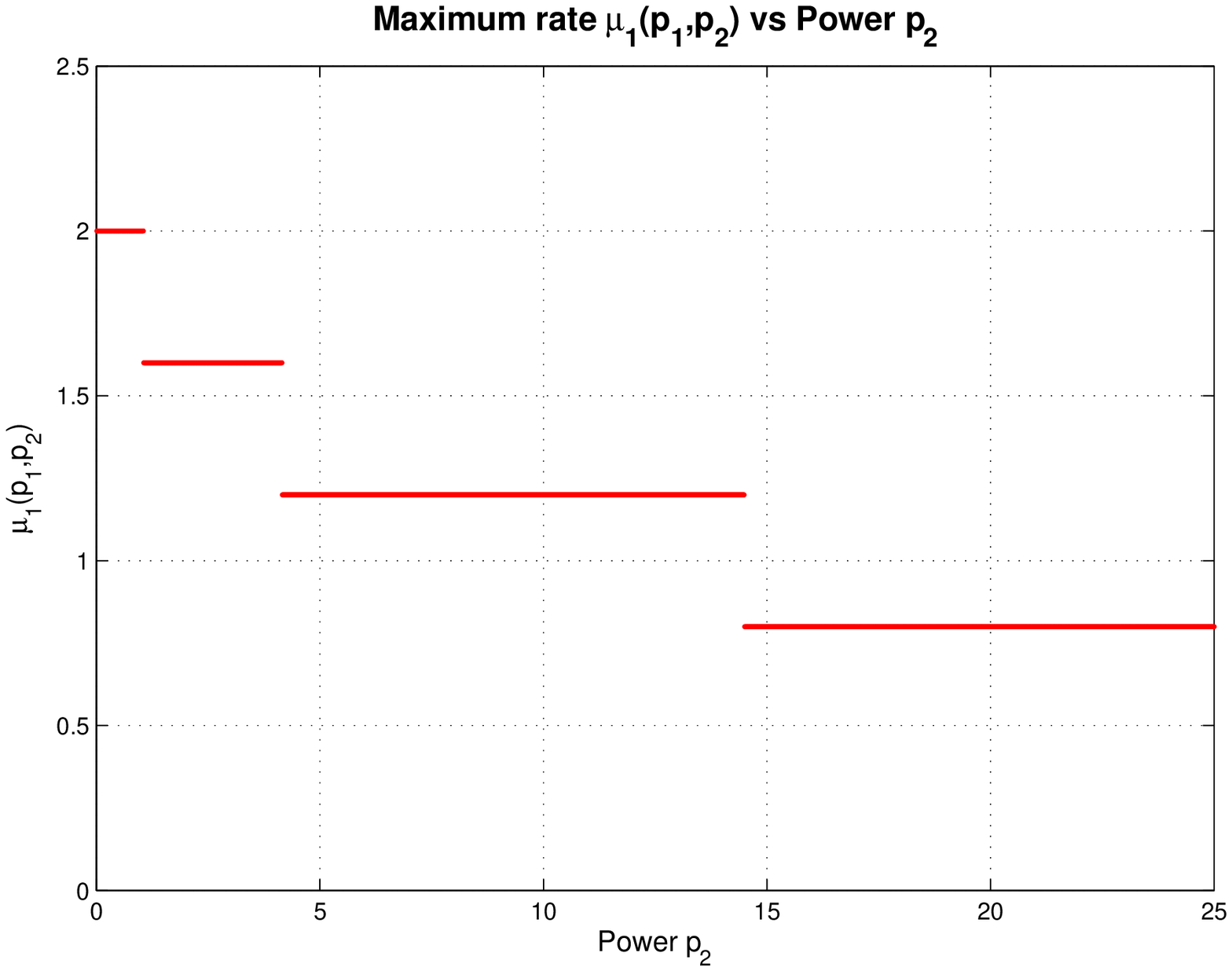}
\caption{We simulate a 2-user Rayleigh/Rayleigh fading channel with set of rates $\mathcal{M} = \left\{0.4,0.8,\ldots,2\right\}$ and $p_2 = 20$ Watt, $P_1 = 25$ Watt. Properties \textbf{P'.2} and \textbf{P'.4} are illustrated a. Maximum Goodput $g_1(p_1,p_2)$ vs power $p_2$, b. Optimal rate $\bar{\mu}_1\left(p_1,p_2\right)$ vs power $p_2$.}
\label{ITW09:fig3}
\end{figure}

\begin{figure}[t]
\centering
\includegraphics[width=9cm]{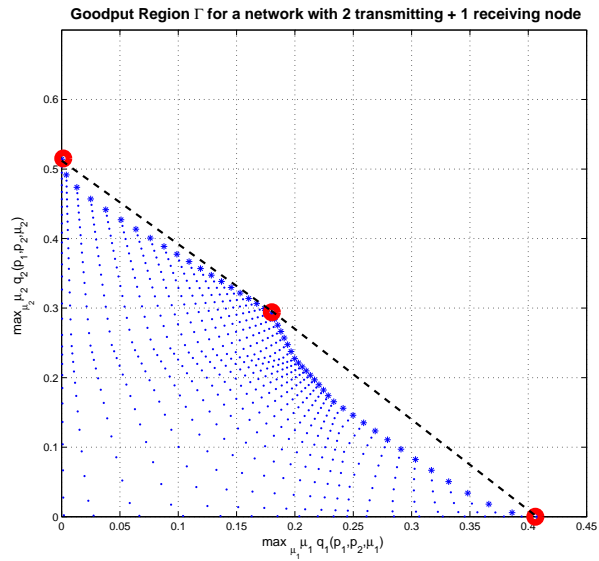}
\caption{The 2-user Rayleigh/Rayleigh goodput region $\hat{\Gamma}_1$ and $\Gamma_1$ for the network of 2 transmitters and a single receiver. The convex hull is shown with the dashed dot lines. For the illustration $\mathcal{M} = \left\{0.4,0.8,\ldots,1.8\right\}$, $P_1^{\max} = 2$ Watt, $P_2^{\max} = 3$ Watt and the success probability function in (\ref{ITW09:eq03}) has been used with $G_{1,1} = G_{1,2} = 1, G_{2,2} = G_{2,1} = 1$.}
\label{ITW09:fig4}
\end{figure}

\begin{figure}[b]
\centering
\includegraphics[width=9cm]{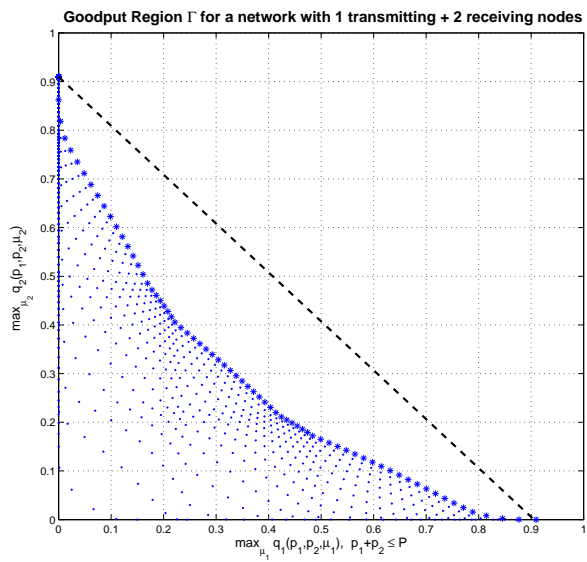}
\caption{The 2-user Rayleigh/Rayleigh goodput region $\hat{\Gamma}_2$ and $\Gamma_2$ for the network consisting of 1 transmitter and 2 receivers. The convex hull is shown with the dashed dot lines. For this topology, $\mathcal{M} = \left\{0.2,0.4,0.6\right\}$, $p_1+p_2\leq P = 10$ Watt and the success probability function in (\ref{ITW09:eq03}) has been used with $G_{1,1} = G_{2,2} = 1, G_{1,2} = 0.5, G_{2,1} = 0.8$.}
\label{ITW09:fig5}
\end{figure}

The above properties are illustrated in Fig.\ref{ITW09:fig2} and Fig.\ref{ITW09:fig3} using a success probability function with the expression in (\ref{ITW09:eq03}) for the 2-user Rayleigh/Rayleigh fading case. These will not be directly used in what follows but are rather useful for the characterisation of the stability region and the optimal scheduling policies of such systems. Examples of the goodput region are shown in figures Fig.\ref{ITW09:fig4} and Fig.\ref{ITW09:fig5} for two simple network topologies: 2 transmitting nodes with 1 receiving, 1 transmitting node with 2 receiving.

\begin{Rem}
In economic terms, we can interpret the success probability function $q_l$ as the \textit{demand} of product $l$ in a market of $L$ firms.  In this framework $\mu_l$ is the \textit{product's price} and $g_l$ is the \textit{firm's revenue}. Then $g_l\left(p_l,\vec{p}_{-l},\mu_l\right) = \mu_l\times q_l\left(p_l,\vec{p}_{-l},\mu_l\right)$ is firm's $l$ (\textit{revenue}) = (\textit{price})$\times$(\textit{demand}). The demand is by \textbf{P.3} a decreasing function of the price, is increasing by \textbf{P.1} in $p_l$ and decreasing by \textbf{P.2} in $p_k,\ k\neq l$. Then $p_l$ can be interpreted as a variable valuating product's $l$ quality (or maybe the money spent by firm $l$ in advertisement) and $p_k$ as the quality (or money for advertisement) of products from competitors $k\neq l$. Then the maximum goodput $g_l\left(\vec{p}\right)$ is the \textit{maximum revenue} that a firm $l$ can obtain by choosing an \textit{optimal price} $\bar{\mu}_l\left(\vec{p}\right)$, given a vector $\vec{p}$. By properties \textbf{P'.1} and \textbf{P'.2} the maximum revenue is increasing in $p_l$ and decreasing in $p_k,\ k\neq l$, whereas by \textbf{P'.3} and \textbf{P'.4} the optimal price is also increasing in $p_l$ and decreasing in $p_k$. Notice that if in \textbf{P.4} log-supermodularity would be replaced by log-submodularity the optimal price would be a decreasing function of $p_l$.
\end{Rem}

\section{NUM Problem Dual Decomposition}
The utility maximization problem in (\ref{ITW09:eq06}) given the network capacity region in Th. \ref{ITW09:Th1} takes the form

\begin{eqnarray}
\label{ITW09:eq15}
\begin{tabular}{l l}
$\mathbf{max}_{x_s\geq 0,\ g_l^d\geq 0}$ & $\sum_{s\in \mathcal{S}}U_s\left(x_s\right)$\\
\textbf{subject to} & $\sum_{l:e(l)=n}g_l^d + x_n^d \leq \sum_{k:b(k)=n}g_k^d$ $\forall n, d$\\
& $\vec{g}\in \mathbf{co}\left(\hat{\Gamma}\right) = \Gamma$
\end{tabular}\nonumber
\end{eqnarray}
and $\hat{\Gamma}$ is given in (\ref{ITW09:eq07}). The constraint set is convex and compact (see \cite[Appendix 4.C]{NeelyThesis}), the objective function is concave and Slater's condition can be shown to hold, hence strong duality also holds and known distributed algorithms, like the one following, can solve the Lagrange dual problem $\min_{\vec{\lambda}\geq 0} L\left(\vec{\lambda}\right)$ which involves the $(N\cdot D)$-vector $\vec{\lambda}$ of dual variables $\lambda_n^d$. The Lagrangian associated with the primal NUM problem is denoted by $L\left(\vec{x}, \vec{\lambda}\right)$ while the dual function $L\left(\vec{\lambda}\right)$ yields, due to the linearity of the differential operator (see \cite{KellyElastic}, \cite{LinShroffCDC04}, \cite{ChenLowChiang06})

\begin{eqnarray}
\label{ITW09:eq17}
L\left(\vec{\lambda}\right) & = &  \sum_{s\in\mathcal{S}}\max_{x_s\geq 0} \left\{U_s\left(x_s\right) - \lambda_s x_s\right\}+ \max_{\vec{g}\in \Gamma}\ \sum_{n,d}\lambda_n^d\left(\sum_{k:b(k)=n}g_k^d - \sum_{l:e(l)=n}g_l^d\right)\nonumber
\end{eqnarray}
and $\lambda_s = \lambda_n^d$, $x_s = x_n^d$ for the flow $s$ with source node $n$ and destination $d$. We can interpret $\lambda_n^d$ as the implicit cost per pair $\left(n,d\right)$. Thus, the NUM problem is decomposed into: 

(a) \textit{The input rate control problem} 

\begin{eqnarray}
\label{ITW09:eq18}
\mathbf{Prob.1:} & & \sum_{s\in\mathcal{S}}\max_{x_s\geq 0} \left\{U_s\left(x_s\right) - \lambda_s x_s\right\}
\end{eqnarray}
solved for each commodity flow at the incoming nodes independently $x_s = U_s\acute{}\ ^{-1}\left(\lambda_s\right)$. Observe that by assumption $U_s\acute{}\left(x_s\right)$ is continuous and monotone decreasing in $\mathbb{R}_+$ (thus a bijection) and the inverse of the function exists. Since $U_s\left(x_s\right)$ is strictly concave the solution is unique for each $\lambda_s$. 

(b) \textit{The scheduling problem}

\begin{eqnarray}
\label{ITW09:eq19}
\mathbf{Prob.2:}& \max_{\vec{g}\in \Gamma}\ \sum_{n,d}\lambda_n^d\left(\sum_{k:b(k)=n}g_k^d - \sum_{l:e(l)=n}g_l^d\right) & = \max_{\vec{g}\in \Gamma}\ \sum_{l,d} g_{l}^d\cdot\left(\lambda_{b(l)}^d-\lambda_{e(l)}^d\right)\nonumber \\
& & \leq \max_{\vec{g}\in \Gamma}\sum_{l}w_l\cdot g_{l}
\end{eqnarray}
Through each link $l$ the commodity $d^* = \arg\max_d \left(\lambda_{b(l)}^d-\lambda_{e(l)}^d\right)$ is scheduled to be routed with goodput rate $g_l$ and $w_l = \max_d \left(\lambda_{b(l)}^d-\lambda_{e(l)}^d,0\right)$. This is the well known \textit{backpressure policy} \cite{TassMultihop}. The solution of (\ref{ITW09:eq19}) further provides the optimal multicommodity goodput flow variables $\left\{g_l^*\right\}$. The optimal solution described is very similar to the DRPC policy in \cite{NeelyRouting}.

If we can solve (\ref{ITW09:eq19}) distributedly, then algorithms can be provided, that solve the dual problem $\min L\left(\vec{\lambda}\right)$ also in a distributed manner, and converge to the optimal average incoming rate vector $\vec{x}^*$ and average price vector $\vec{\lambda}^*$. The dual problem can be solved by the subgradient method. The prices $\lambda_n^d$ for each node-destination pair $\left(n,d\right)$ are step-wise adjusted by

\begin{eqnarray}
\label{ITW09:eq22}
\lambda_n^d\left(t+1\right) & = & [\lambda_n^d\left(t\right) + \gamma_t\cdot (x_n^d\left(t\right)- \sum_{k:b(k)=n}g_k^d\left(t\right) + \sum_{l:e(l)=n}g_l^d\left(t\right) )]^+.
\end{eqnarray}
In the above $\gamma_t$ is a positive scalar stepsize, $\left[\ldots\right]^+$ denotes the projection onto the set $\mathbb{R}_+$ and for each $t$ the values $x_n^d\left(t\right)$ and $g_n^d\left(t\right)$ are calculated by solving problems (\ref{ITW09:eq18}) and (\ref{ITW09:eq19}) respectively and using prices $\vec{\lambda}\left(t\right)$. As noted in the aforementioned works, in practice a constant stepsize is used for implementation purposes, although the convergence of the algorithm is guaranteed for $\gamma_t\stackrel{t\rightarrow\infty}{\rightarrow}0$. For constant stepsizes statistical convergence to $\vec{\lambda}^*,\vec{x}^*$ is guaranteed as shown in \cite[Def.1,Th.2]{ChenLowChiang06}.

\section{The Scheduling Problem}

\subsection{Relaxation}
As was mentioned previously it is very important that the problem in (\ref{ITW09:eq19}) is solved in a \textit{distributed} manner. To this aim the theory of \textit{supermodular games} can be used. We make the following two assumptions

\begin{enumerate}
\item \textit{Assumption 1}: Each origin node chooses a \textit{single} end node to transmit
\item \textit{Assumption 2}: Each node can transmit and receive at the same time
\item \textit{Assumption 3}: Fixed scheduled rates per link $\mu_l$ are considered.
\end{enumerate}
Specifically the last assumption constraints the generality of the initial model but is necessary for the approach that follows. Variable scheduled rates would involve a joint maximization over power allocation and rates. This would complicate the analysis, but is a rather important topic for future research. The maximization in (\ref{ITW09:eq19}) can be written as

\begin{eqnarray}
\max_{\vec{g}\in \Gamma}\sum_{l}w_l \cdot g_{l} \stackrel{(a)}{=}
\max_{\vec{g}\in \hat{\Gamma}}\sum_{l}w_l \cdot g_{l} & = &
\max_{\vec{p}\in\Pi }\sum_{n:n=b\left(l\right)}\sum_{n:n=e\left(l\right)}w_l \cdot \mu_l q_{l}\left(\vec{p},\mu_l\right) \nonumber\\
\label{ITW09:eq23}
& \stackrel{(b)}{=} & \max_{p_n\in\left[P_n^{\min},P_n^{\max}\right]}\sum_{n:n=b\left(l\right)}w_n \cdot \mu_n q_{n}\left(\vec{p},\mu_n\right)
\end{eqnarray}
where (a) comes from the fact that the objective function is linear and the supporting hyperplanes to the sets $\hat{\Gamma}$ and $\Gamma = \mathbf{co}\left\{\hat{\Gamma}\right\}$ are the same, while (b) from Assumption 1. The latter simplifies the problem to a weighted sum maximization problem with number of summants equal to the number of nodes and allows the formulation of a noncooperative game in the following subsections, where each node decides independently over its transmitting power through a single chosen link. The capacity region of the system in Theorem \ref{ITW09:Th1} is of course reduced. An important question is how each node choses the optimal single link to transmit.

The number of links with node $n$ as origin are all links $\left\{l\in\mathcal{L}: n=b\left(l\right)\ \& \ w_l>0\right\}$. This is the \textit{connectivity set} of node $n$. The optimal link is obviously the one which provides maximum weighted goodput to the above summation, for a given power allocation $p_n\in\left[P_n^{\min}, P_n^{\max}\right]$. 

Departing here briefly from the main line of the analysis, we end this subsection with a heuristic suggestion for an almost optimal choice of a single receiver node, using low information exchange between the network nodes. Applying Markov's inequality in (\ref{ITW09:eq01})

\begin{eqnarray}
\label{ITW09:eq24a}
\omega_l \mu_l \mathbb{P}\left(SINR_l\left(\vec{p}\right)\geq e^{\mu_l}-1\right)\leq\frac{\omega_l \mu_l} {e^{\mu_l}-1}\mathbb{E}\left(SINR_l\left(\vec{p}\right)\right). \nonumber
\end{eqnarray}
Suppose that the end node $e\left(l\right)$ of each link $l$ measures the received level of interference, the latter denoted by $\mathcal{I}_l$. This is not any more a random variable with unknown realization but rather a known deterministic quantity. The right handside then reduces to

\begin{eqnarray}
\label{ITW09:eq24}
\frac{\omega_l \mu_l}{e^{\mu_l}-1}\frac{\mathbb{E}\left(G_{l,l}F_{l,l}p_l\right)}{\left(\mathcal{I}_l + \sigma^2_{e\left(l\right)}\right)}& = & 
\frac{\omega_l \mu_l }{e^{\mu_l}-1} \frac{G_{l,l}p_l}{\left(\mathcal{I}_l + \sigma^2_{e\left(l\right)}\right)}.
\end{eqnarray}
This is an upper bound on the actual error probability. The process of the sub-optimal choice then is as follows. Each destination node of links belonging to the connectivity set of node $n$, informs the origin node over $\mathcal{I}_l+\sigma^2_{e\left(l\right)}$ and afterwards $n$ chooses to transmit over the link with the maximum ratio (\ref{ITW09:eq24}), since $G_{l,l}$, $\omega_l$ and $\mu_l$ are known to $n$. 

An alternative way to choose a single receiver node could be by assigning to each element of the connectivity set a probability, with sum equal to one per transmitting node, and the choice will then be a random process.

\subsection{Optimality conditions}
Using the Karush-Kuhn-Tucker (KKT) optimality conditions and observing that the active inequality constraint gradients are linearly independent \cite[pp.315-317]{BertsekasNP} all feasible vectors $\vec{p}$ are regular and we have the following \textit{necessary conditions} for $\vec{p}^*$ to be a local maximizer of the objective function in (\ref{ITW09:eq23}).

\begin{eqnarray}
\label{ITW09:eq25}
0 = \frac{\partial}{\partial p_n}\mathbf{L}\left(\vec{p}^*,\vec{\nu}^l,\vec{\nu}^u\right) = \nu^l_n - \nu^u_n + \omega_n\cdot \mu_n\frac{\partial q_n\left(\vec{p}^*,\mu_n\right)}{\partial p_n} + \sum_{m\neq n}\omega_m\cdot \mu_m\frac{\partial q_m\left(\vec{p}^*,\mu_m\right)}{\partial p_n}
\end{eqnarray}
for each $n$ and the complementary slackness conditions satisfy
\begin{eqnarray}
\label{ITW09:eq26}
\nu_n^l\cdot \left(P^{\min}_n-p^*_n\right) = 0 \ \ \ \& & \nu_n^u\cdot \left(p^*_n-P^{\max}_n\right) = 0
\end{eqnarray}
$\mathbf{L}\left(\vec{p}^*,\vec{\nu}^l,\vec{\nu}^u\right)$ is the Lagrangian of the problem in (\ref{ITW09:eq23}). The conditions are only necessary and \textbf{not} sufficient. We make here the remark that if the objective function were concave, the dual gap would be zero and any local maximizer would also be global for the problem at hand. In this case the conditions would also be sufficient. Unfortunately this generaly does not hold for the specific objective function.

Divide (\ref{ITW09:eq25}) and (\ref{ITW09:eq26}) by $q_n\left(\vec{p}^*,\mu_n\right)$ (which is definitely positive if we choose $P_n^{\min}>0,\ \forall n\in \mathcal{N}$) and then - approching the problem similarly to \cite{HuangBerryHonigGame} - set

\begin{eqnarray}
\label{ITW09:eq27a}
0\geq \omega_m \cdot \mu_m \frac{\partial q_m\left(\vec{p},\mu_m\right)}{\partial p_n}\frac{1}{q_n\left(\vec{p},\mu_n\right)} & = & \frac{\pi_{m,n}\left(\vec{p}\right)}{q_n\left(\vec{p},\mu_n\right)} \\
\label{ITW09:eq27}
& = & \hat{\pi}_{m,n}\left(\vec{p}\right)
\end{eqnarray}
and for the Lagrange multipliers
\begin{eqnarray}
\label{ITW09:eq27b}
\hat{\nu}_n^u = \frac{\nu_n^u}{q_n\left(\vec{p}^*,\mu_n\right)}\geq 0 & \&\ \ \hat{\nu}_n^l = \frac{\nu_n^l}{q_n\left(\vec{p}^*,\mu_n\right)}\geq 0.
\end{eqnarray}
With the above substitutions the per node condition in (\ref{ITW09:eq25}) is rewritten as
\begin{eqnarray}
\label{ITW09:eq28}
\omega_n\cdot \mu_n \frac{1}{q_n\left(\vec{p}^*,\mu_n\right)}\frac{\partial q_n\left(\vec{p}^*,\mu_n\right)}{\partial p_n}+
\sum_{m\neq n}\hat{\pi}_{m,n}\left(\vec{p}^*\right) & =
\hat{\nu}^u_n - \hat{\nu}^l_n.
\end{eqnarray}
Then (\ref{ITW09:eq28}) with the related complementary slackness conditions are the necessary and sufficient conditions for $p_n^*$ to be the global maximizer of the problem

\begin{eqnarray}
\label{ITW09:eq29}
\mathbf{\max}_{p_n} & \omega_n \mu_n \log\left(q_n\left(p_n;\vec{p}^*_{-n},\mu_n\right)\right)+ p_n\sum_{m\neq n}\hat{\pi}_{m,n}\left(\vec{p}^*\right)
\end{eqnarray}
since by property \textbf{P.1} of the success function, $\log q_n\left(\vec{p},\mu_n\right)$ is concave in $p_n$, the constraint set $p_n\in{\left[P_n^{\min},P_n^{\max}\right]}$ is convex and compact and Slater's condition holds true. This explains now why the division in (\ref{ITW09:eq27a}) and (\ref{ITW09:eq27b}) was required.

\subsection{A Supermodular Game}
If we view $-\hat{\pi}_{m,n}\left(\vec{p}\right)$ as the price charged by user $m$ to user $n$ for affecting its goodput by creating interference, we can approach the solution to the optimal power allocation problem in a distributed fashion with the use of game theory. 
We denote the noncooperative game by the triple $\mathcal{G} = \left(\mathcal{N},\Pi,\left\{J_n\left(\cdot\right),\ n\in\mathcal{N}\right\}\right)$ where $\mathcal{N}$ are the players, $\Pi$ is the set of feasible joint strategies and $J_n$ is the payoff function for user $n$. 

We distinguish between two types of players. First, the \textit{power players} who belong to the set $\mathcal{N}^p$, each one of which represents a node and the set of feasible joint strategies $\Pi^p$ is identical to the set $\Pi$ of feasible power allocations. Their payoff function equals

\begin{eqnarray}
\label{ITW09:eq30}
J_n\left(p_n;\vec{p}_{-n},(\hat{\pi}_{m,n})\right) & = & \omega_n \mu_n \log\left(q_n\left(p_n;\vec{p}_{-n},\mu_n\right)\right)+ p_n \cdot \sum_{m\neq n}\hat{\pi}_{m,n}
\end{eqnarray}
We often set $c_n := \sum_{m\neq n}\hat{\pi}_{m,n}$ to emphasize the dependence of $J_n$ on the sum instead of the individual prices. The \textit{best response correspondence for player $n$} is the set

\begin{eqnarray}
\label{ITW09:eq31}
Y_n\left(\vec{p}_{-n}\right) = \arg \max_{p_n\in \Pi_n\left(\vec{p}_{-n}\right)} J_n\left(p_n;\vec{p}_{-n}, (\hat{\pi}_{m,n})\right)
\end{eqnarray}
where $\Pi_n\left(\vec{p}_{-n}\right) = \left\{p_n:\left(p_n,\vec{p}_{-n}\right)\in \Pi^p\right\}$. 

Second, the \textit{price players} who belong to the set $\mathcal{N}^{pr}: = \left\{\left(m,n\right):m\neq n, \ m,n\in\mathcal{N}\right\}$ with cardinality $N\times (N-1)$. The feasible set of strategies for player $\left(m,n\right)$ is

\begin{eqnarray}
\label{ITW09:eq33}
\Pi_{m,n}^{pr} = \left\{\hat{\pi}_{m,n}\in\left[\min_{\vec{p}\in\Pi}\hat{\pi}_{m,n}\left(\vec{p}\right),0\right]\right\}
\end{eqnarray}
where $\hat{\pi}_{m,n}\left(\vec{p}\right)$ is given in (\ref{ITW09:eq27}). The best response for a price player is denoted by (following \cite{HuangBerryHonigGame})

\begin{eqnarray}
\label{ITW09:eq34}
Y_{m,n}^{pr} = \arg \max_{\hat{\pi}_{m,n}\in \Pi_{m,n}^{pr}} - \left(\hat{\pi}_{m,n}-\hat{\pi}_{m,n}\left(\vec{p}\right)\right)^2
\end{eqnarray}
and $\Pi^{pr} = \left\{\Pi_{\left(2,1\right)},\ldots, \Pi_{\left(N-1,N\right)}\right\}$ is the joint feasible set.

A \textit{Nash equilibrium} (NE) for the game $\mathcal{G}$ is defined as the set of power vectors $\vec{p}^e = \left(p_1^e,\ldots,p_N^e\right)$ and price vectors $\vec{\hat{\pi}^e} = \left(\hat{\pi}^e_{2,1},\ldots,\hat{\pi}^e_{n,1},\ldots,\hat{\pi}^e_{1,n},\ldots,\hat{\pi}^e_{n-1,n}\right)$ with the property for every $n,m\in\mathcal{N}^p$ and every $\left(m,n\right)\in\mathcal{N}^{pr}$

\begin{eqnarray}
\label{ITW09:eq32}
J_n\left(p_n^e,\vec{p}_{-n}^e,(\hat{\pi}_{m,n}^e)\right)\geq J_n\left(p_n,\vec{p}_{-n}^e,(\hat{\pi}_{m,n}^e)\right), & \& &
\hat{\pi}_{m,n}^e = \hat{\pi}_{m,n}\left(\vec{p}^e\right), \ \forall p_n\in \Pi_n\left(\vec{p}_{-n}^e\right)
\end{eqnarray}
Hence $p_n^e$ belongs to the best response correspondence of player $n$, $\forall n\in\mathcal{N}^p$, given the equilibrium prices, whereas $\hat{\pi}^e_{m,n}$ belongs to the best response correspondence of player $\left(m,n\right)\in\mathcal{N}^{pr}$ given the equilibrium powers. 

The existence and uniqueness of the NE when the prices do not take part as players in the game has been proven in \cite[Th.III.1]{AlpcanBasarOutTrans} under mild assumptions on the problem parameters usually satisfied in practice. In our case however with $N+N\times \left(N-1\right) = N^2$ players the uniqueness of a Nash equilibrium is not guaranteed. We can however make use of the theory of \textit{supermodular games}, exploiting the structure of the payoff function in (\ref{ITW09:eq30}) to find algorithms that converge to one of the Nash Equilibria. We first give the definition of a supermodular game from Topkis \cite{Topkis}

\begin{Def}
\label{ITW09:Def2}
A noncooperative game with $N$ players $\left\{\mathcal{N},\Pi,\left\{f_n:n\in\mathcal{N}\right\}\right\}$, each having strategy $x_n$ belonging to the feasible set of strategies $\Pi_n\left(\vec{x}_{-n}\right)$, is supermodular if the set $\Pi$ of feasible joint strategies is a sublattice of $\mathbb{R}^N$ and for each $n$ the payoff function $f_n$ is supermodular in player $n$'s strategy $x_n$ and has increasing differences for all pairs $\left(x_i,x_j\right)\in \Pi_i\times \Pi_j$, $i\neq j$, $i,j\in \mathcal{N}$.
\end{Def}

\begin{Theorem}
\label{ITW09:Th5}
The noncooperative game with $N$ power players and $N\times (N-1)$ price players is a supermodular game \cite[p.178]{Topkis}. Furthermore, the set of equilibrium points is a nonempty complete lattice and a greatest and least equilibrium point exist.
\end{Theorem}

\begin{proof}
See Appendix B.
\end{proof}
After proving that the problem at hand has the desired properties so that supermodular game theory can be applied we prove in the following that the Nash Equilibria of the game are exactly the power allocations that satisfy the KKT necessary optimality conditions of the original sum weighted maximization problem.
\begin{Theorem}
\label{ITW09:Th6}
Under the condition that $\forall n$, $P_n^{\min}>0$, a power vector $\vec{p}^e$ is a Nash Equilibrium of the supermodular power-price game if and only if it satisfies the necessary optimality conditions (\ref{ITW09:eq25})-(\ref{ITW09:eq26}).
\end{Theorem}

\begin{proof}
See Appendix C.
\end{proof}

The above theorem is rather important because it shows that the formulated game leads to one of the solutions of the scheduling problem. If the objective function in (\ref{ITW09:eq23}) is concave then the NE is also unique and the game converges to the unique global maximizer. The suboptimality of the proposed scheme in the current work thus lies solely on the fact that the KKT conditions are only necessary but not sufficient. If we can define the region of $\Pi$ for which the objective function is concave and restrict the feasible power allocations to that, the suggested distributed solution is the optimal one. This can be a topic for future investigations.

\subsection{The Scheduling Algorithm}
\label{ITW09:SecC}
In the current paragraph we provide an algorithm which updates for each player the power allocation $p_n$ and the price $\pi_{m,n}$. Starting from any initial point within the joint feasible region, the algorithm will eventually converge to a NE bounded component-wise by the greatest and least NE. It is related to the Round-Robin optimization for supermodular games \cite[Ch. 4.3.1]{Topkis}, versions of which are suggested in \cite{HuangBerryHonigGame} and \cite{PowerPriceSMG02}.

The algorithm has two phases for each iteration $t$ and is given in Table I. The \textit{power update phase} calculates the best response for each user $n$ by (\ref{ITW09:eq30}) given fixed prices $\hat{\pi}_{m,n}^{(t)}$ and the opponents' decisions $\vec{p}_{-n}^{(t)}$.

During the \textit{price update phase} each user $m$ calculates $(N-1)$ new prices $\pi_{m,n}^{(t)}$ (without the hat)  by (\ref{ITW09:eq27a}) given the updated power vector. Then all users $m\neq n$ communicate the values $\pi_{m,n}^{(t)}$ to user $n$, who divides their sum by $q_n\left(\vec{p},\mu_n\right)$ to form the new sum price $c_n^{(t+1)}$ for the next power update phase.

Observe that for each iteration, user $n$ should know: (a) Its own rate of transmission $\mu_n$ (which defines $q_n$) and weight $\omega_n$, (b) the power profile of the other users $\vec{p}_{-n}$, (c) the prices $\pi_{m,n}$ communicated by the interfering users and (d) the slow fading coefficients $G_{m,n}$ which depend on the distance between the nodes.

\subsection{Implementation Issues}

Considering implementation issues of the algorithm, information (b) and (c) should be communicated to node $n$, while (d) should be globally known. Notice that communicating the information over the power profile of the interfering users will violate the distributed nature of the algorithm. Instead of the power vector $\vec{p}_{-n}$ however, it suffices for each user to measure the current level of interference $\mathcal{I}_n = \sum_{m\neq n} G_{mn}F_{mn}p_m$ in which case we write 

\begin{eqnarray}
\label{ITW09:eq37}
\hat{q}_n\left(p_n,\mathcal{I}_n,\mu_n\right) & = & \mathbb{P}\left(\frac{G_{nn}F_{nn}p_n}{\mathcal{I}_n+\sigma^2_{e\left(l\right)}}\geq \gamma_n\left(\mu_n\right)\right)\stackrel{Rayl.}{=}  \exp\left(\frac{-\left(\mathcal{I}_n+\sigma^2_{e\left(l\right)}\right)\gamma_n\left(\mu_n\right)}{G_{nn}p_n}\right)\nonumber
\end{eqnarray}
where $l:b(l)=n$ and the second equality holds for Rayleigh fading. The payoff function will change accordingly.
In the price update phase observe that the partial derivative of $\hat{q}_m$ with respect to $p_n$ will be given by
\begin{eqnarray}
\label{ITW09:eq39}
\frac{\pi_{m,n}}{\omega_m \mu_m} & = & \frac{\partial \hat{q}_m\left(p_m,\mathcal{I}_m,\mu_m\right)}{\partial p_n}= \frac{\partial \hat{q}_m\left(p_m,\mathcal{I}_m,\mu_m\right)}{\partial \mathcal{I}_m}\frac{\partial \mathcal{I}_m}{\partial p_n}
= -\frac{\phi_{m}\left(p_m,\mathcal{I}_m\right)}{\omega_m \mu_m}G_{nm}F_{nm}
\end{eqnarray}
The new values $\phi_m$ can be computed by each user $m$ and are independent of the destination user $n$.

\begin{eqnarray}
\label{ITW09:eq40}
\phi_{m}\left(p_m,\mathcal{I}_m\right) \stackrel{Rayl.}{=} \omega_m \mu_m\hat{q}_m\left(p_m,\mathcal{I}_m,\mu_m\right)\left(\frac{\gamma_m\left(\mu_m\right)}{G_{mm}p_m}\right)\nonumber
\end{eqnarray}
In the form (\ref{ITW09:eq39}) observe that the actual realization of the random variable $F_{nm}$ appears. Remember that $F_{nm}$ is the fast fading channel power coefficient. This information is \textit{unknown}. But node $n$ is interested in the sum $c_n$ of the prices $\hat{\pi}_{m,n}$ (see (\ref{ITW09:AlgoSum})) which can be written as

\begin{eqnarray}
\label{ITW09:eq41}
c_n = -\frac{1}{q_n\left(p_n;\mathcal{I}_n,\mu_n\right)}\sum_{m\neq n}G_{nm}F_{nm} \phi_{m}\left(p_m,\mathcal{I}_m\right)
\end{eqnarray}
If each node $m\neq n$ \textit{broadcasts} a sequence of random symbols $S_m$, $\left|S_m\right|^2 = 1$ with power $\sqrt{\phi_{m}\left(p_m,\mathcal{I}_m\right)}$ the received signal at node $n$ will be (assuming reciprocity of the channel gains)

\begin{eqnarray}
\label{ITW09:eq42}
Y_n = \sum_{m\neq n}H_{nm}\sqrt{\phi_{m}\left(p_m,\mathcal{I}_m\right)}S_m + Noise
\end{eqnarray}
and its power is $\left|Y_n\right|^2 = \sum_{m\neq n}G_{nm}F_{nm} \phi_{m}\left(p_m,\mathcal{I}_m\right) + \sigma^2_n$. If the receiving node $n$ divides by $-q_n\left(p_n;\mathcal{I}_n,\mu_n\right)$ we get a noisy version of the expression in (\ref{ITW09:eq41}).

The above idea is borrowed from recent works that deal with ways to use the Wireless Multiple Access Channel (MAC) in order to compute general functions of data among which is also addition \cite{NazerGastparInfoTh}. The above method using power to convey information can be found specifically in \cite{MarioWCNC09}. From the above we realize that although the fast fading coefficients are not known to the users $m$ that have to calculate the prices $\pi_{m,n}$ these can be revealed to the receiver $n$ within the sum signal in (\ref{ITW09:eq41}). 

Finally rather important is the fact that for the implementation of the algorithm, each user $m$ has to be aware of \textit{its received interference $\mathcal{I}_m$} and actually calculate only a \textit{single price $\phi_m$}. Then \textit{in a single step} during the price update phase each player/node broadcasts its price $\phi_m$, while acting simultaneously as a receiver (remember Assumption 2) to obtain the channel-power-weighted sum of the prices of the other $N-1$ users. The entire network topology is not any more necessary to be known to each user $m$, \textit{only the slow fading gain $G_{mm}$}. This allows the scheduling algorithm to have as well application in cases where the topology possibly changes due to user mobility.

\section{Simulations}
\begin{figure}[h]
\centering
\includegraphics[width=19cm,height=13cm]{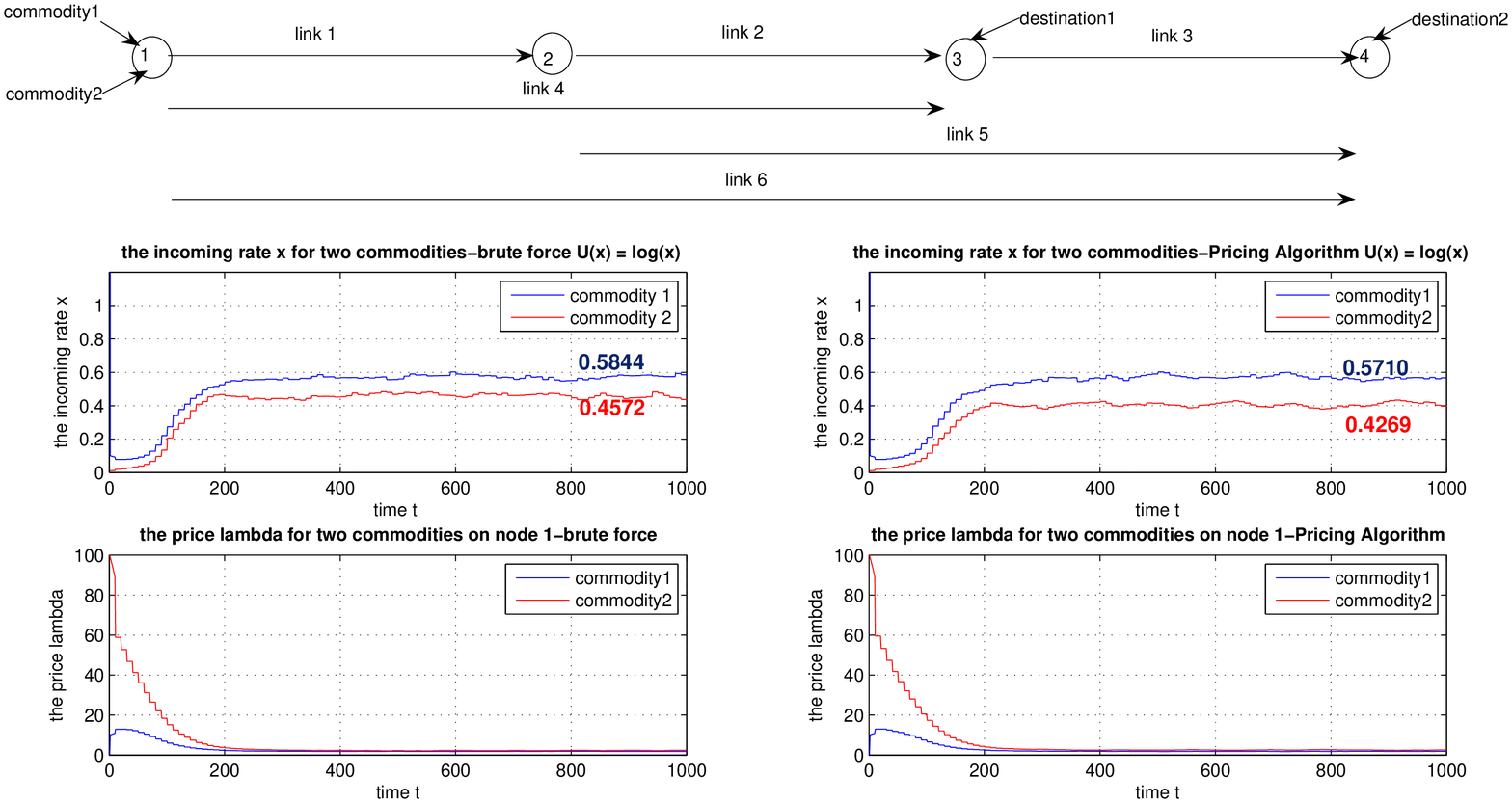}
\caption{Congestion control for a four node topology with two commodity flows. The scheduling problem is solved using the suggested pricing algorithm. Comparison plots with a brut force search to find the global optimum of the weighted sum maximization problem in (\ref{ITW09:eq23}) are provided in the first column.}
\label{ITW08:fig1}
\end{figure}

Simulation results of the proposed scheme for congestion control, routing and distributed power allocation when hop-by-hop retransmissions are taken into account are presented in Fig.\ref{ITW08:fig1}. We used a four node topology having two commodity flows with sorce node 1 and destination nodes 3 and 4 respectively. The congestion control requires the solution of the subproblems (\ref{ITW09:eq18}) and (\ref{ITW09:eq19}) respectively with prices $\vec{\lambda}\left(t\right)$. The prices are updated per node using the expression in (\ref{ITW09:eq22}). The optimal links per node are chosen at each step using (\ref{ITW09:eq24}). The scheduling problem in (\ref{ITW09:eq23}) is solved initially by brut force (left column) to provide a comparison with the results obtained when the price based algorithm is used (right column). We notice that although the uniqueness of the Nash Equilibrium cannot be guaranteed the results of the suggested algorithm considering the maximum supported incoming rate as well as the queue length (price $\lambda_1$) are almost optimal. An important remark is that the two solutions would be exactly the same if the objective function in (\ref{ITW09:eq23}) would be concave.

\section{Appendix}

\subsection{Proof of Theorem \ref{ITW09:Th3}}
In the following we will neglect the dependence of functions on variables that are considered constant throughout a proof.
\begin{itemize}
\item \textbf{P'.1}: Suppose $p_l^+>p_l$ and let $\mu_l^+ :=\arg\max_{\mu_l\in\mathcal{M}}\mu_l q_l\left(p_l^+,\mu_l\right)$, and also $\bar{\mu}_l :=\arg\max_{\mu_l\in\mathcal{M}}\mu_l q_l\left(p_l,\mu_l\right)$. Then $\forall \mu_l\in\mathcal{M}$

\begin{eqnarray}
\label{ITW09:AppIV01}
\mu_l^+ q_l\left(p_l^+,\mu_l^+\right) \stackrel{(a)}{\geq} \mu_l q_l\left(p_l^+,\mu_l\right) \stackrel{(b)}{>} \mu_l q_l\left(p_l,\mu_l\right)\nonumber
\end{eqnarray}
In the above, (a) comes from the definition of $\mu_l^+$ and (b) from \textbf{P.1} of the success probability function. Since the inequality holds $\forall \mu_l$ it also holds for $\mu_l=\bar{\mu}_l$, hence $g_l\left(p_l^+,\vec{p}_{-l}\right)> g_l\left(p_l,\vec{p}_{-l}\right)$.

\item \textbf{P'.2}: For the monotonicity we proceed as above, where $p_k^+>p_k$, $\mu_l^+ :=\arg\max_{\mu_l\in\mathcal{M}}\mu_l q_l\left(p_k^+,\mu_l\right)$, and also $\bar{\mu}_l :=\arg\max_{\mu_l\in\mathcal{M}}\mu_l q_l\left(p_k,\mu_l\right)$. Then $\forall \mu_l\in\mathcal{M}$

\begin{eqnarray}
\label{ITW09:AppIV02}
\bar{\mu}_l q_l\left(p_k,\bar{\mu}_l\right) \stackrel{(c)}{\geq} \mu_l q_l\left(p_k,\mu_l\right) \stackrel{(d)}{>} \mu_l q_l\left(p_k^+,\mu_l\right)\nonumber
\end{eqnarray}
where (c) comes from the definition of $\mu_l^+$ and (d) from the monotonicity in \textbf{P.2}. Since the inequality holds $\forall \mu_l$ it also holds for $\mu_l=\mu_l^+$, hence $g_l\left(p_k^+,\vec{p}_{-k}\right)< g_l\left(p_k,\vec{p}_{-k}\right)$.

For the convexity we write for $p_k^{(1)}\neq p_k^{(2)}$
\end{itemize}

\begin{eqnarray}
\label{ITW09:AppIV03}
g_l\left(\theta p_k^{(1)} + \left(1-\theta\right)p_k^{(2)}\right) & = &\max_{\mu_l}\mu_l q_l\left(\theta p_k^{(1)} + \left(1-\theta\right)p_k^{(2)},\mu_l\right)\stackrel{\textbf{(P.2)}}{\leq}\nonumber\\
\max_{\mu_l}\left\{\theta \mu_l q_l\left(p_k^{(1)},\mu_l\right)+\left(1-\theta\right)\mu_l q_l\left(p_k^{(2)},\mu_l\right)\right\}
& \leq & \max_{\mu_l}\theta \mu_l q_l\left(p_k^{(1)},\mu_l\right)+\max_{\mu_l}\left(1-\theta\right)\mu_l q_l\left(p_k^{(2)},\mu_l\right)\nonumber\\
& = & \theta g_l\left(p_k^{(1)},\vec{p}_{-k}\right) + \left(1-\theta\right)g_l\left(p_k^{(2)},\vec{p}_{-k}\right)\nonumber
\end{eqnarray}

\begin{itemize}
\item \textbf{P'.3} Choose $p_l^b\geq p_l^a$ and denote $\mu_l^b :=\arg\max_{\mu_l\in\mathcal{M}}\mu_l q_l\left(p_l^b,\mu_l\right)$, and also $\mu_l^a :=\arg\max_{\mu_l\in\mathcal{M}}\mu_l q_l\left(p_l^a,\mu_l\right)$. By definition 

\begin{eqnarray}
\label{ITW09:AppIV04}
\mu_l^b q_l\left(p_l^b,\mu_l^b\right) \geq \mu_l^a q_l\left(p_l^b,\mu_l^a\right) & \Rightarrow & 
\frac{\mu_l^b}{\mu_l^a} \geq \frac{q_l\left(p_l^b,\mu_l^a\right)}{q_l\left(p_l^b,\mu_l^b\right)}
\end{eqnarray}
We prove the property by contradiction. Suppose that $\mu_l^b < \mu_l^a$. 
From the log-supermodularity property \textbf{P.4}

\begin{eqnarray}
\label{ITW09:AppIV05}
\frac{q_l\left(p_l^b,\mu_l^a\right)}{q_l\left(p_l^a,\mu_l^a\right)} > \frac{q_l\left(p_l^b,\mu_l^b\right)}{q_l\left(p_l^a,\mu_l^b\right)}
\end{eqnarray}
Combining (\ref{ITW09:AppIV04}) and (\ref{ITW09:AppIV05})

\begin{eqnarray}
\label{ITW09:AppIV06}
\frac{\mu_l^b}{\mu_l^a} \stackrel{(\ref{ITW09:AppIV04})}{\geq} \frac{q_l\left(p_l^b,\mu_l^a\right)}{q_l\left(p_l^b,\mu_l^b\right)} \stackrel{(\ref{ITW09:AppIV05})}{>} \frac{q_l\left(p_l^a,\mu_l^a\right)}{q_l\left(p_l^a,\mu_l^b\right)} & \Rightarrow &
\mu_l^b q_l\left(p_l^a,\mu_l^b\right) > \mu_l^a q_l\left(p_l^a,\mu_l^a\right)
\end{eqnarray}
But (\ref{ITW09:AppIV06}) is impossible from the definition of $\mu_l^a :=\arg\max_{\mu_l\in\mathcal{M}}\mu_l q_l\left(p_l^a,\mu_l\right)$ hence $\mu_l^b\geq \mu_l^a$.

\item \textbf{P'.4}: We make use of the fact that given a pair $\left(p_l^{(1)},\mu_l^{(1)}\right)$ there always exists another one $\left(p_l^{(2)},\mu_l^{(2)}\right)$, with $p_l^{(1)}\neq p_l^{(2)}$ and $\mu_l^{(1)}\neq \mu_l^{(2)}$ such that $q_l\left(p_l^{(1)},\vec{p}_{-l},\mu_l^{(1)}\right) = q_l\left(p_l^{(2)},\vec{p}_{-l},\mu_l^{(2)}\right)$. This is because $\forall \mu_l\in \mathcal{M}$, the success probability function $q_l\left(p_l,\mu_l\right)\in\left[0,1\right]$ is strictly increasing in $p_l$ and strictly decreasing in $\mu_l$ by \textbf{P.1} and \textbf{P.3} (here $p_l\in\mathbb{R}_+$).

Denote by $\mu_l^b :=\arg\max_{\mu_l\in\mathcal{M}}\mu_l q_l\left(p_k^b,\mu_l\right)$, and also $\mu_l^a :=\arg\max_{\mu_l\in\mathcal{M}}\mu_l q_l\left(p_k^a,\mu_l\right)$.

Using the above fact we can write $q_l\left(p_l,p_k^a,\mu_l^a\right) = q_l\left(p_l^a,p_k^a,\mu_l\right)$ and $q_l\left(p_l,p_k^b,\mu_l^b\right) = q_l\left(p_l^b,p_k^b,\mu_l\right)$, e.g. for some $\mu_l\geq \max\left\{ \mu_l^a,\mu_l^b\right\}$, $p_l^a\geq p_l$ and $p_l^b\geq p_l$. By definition

\begin{eqnarray}
\label{ITW09:AppIV07}
\frac{\mu_l^b}{\mu_l^a} \geq \frac{q_l\left(p_l,p_k^b,\mu_l^a\right)}{q_l\left(p_l,p_k^b,\mu_l^b\right)} = \frac{q_l\left(p_l^a,p_k^b, \mu_l\right)}{q_l\left(p_l^b, p_k^b, \mu_l\right)}
\end{eqnarray}

The property is proven by contradiction.. Choose $p_k^b\geq p_k^a$ and suppose that 

\begin{eqnarray}
\label{ITW09:AppIV08}
\mu_l^b \geq \mu_l^a & \stackrel{(\textbf{P.3})}{\Leftrightarrow} &
q_l\left(p_l,p_k^b,\mu_l^b\right) \leq q_l\left(p_l,p_k^b,\mu_l^a\right)\Leftrightarrow\nonumber\\
q_l\left(p_l^b,p_k^b,\mu_l\right) \leq q_l\left(p_l^a,p_k^b,\mu_l\right) & \stackrel{(\textbf{P.1})}{\Leftrightarrow} & 
p_l^b \leq p_l^a
\end{eqnarray}

From the log-supermodularity property \textbf{P.5} we reach the inequality $\mu_l^b q_l\left(p_l, p_k^a, \mu_l^b\right)\geq \mu_l^a q_l\left(p_l, p_k^a, \mu_l^a\right)$ which is impossible by the definition of $\mu_l^a$ $\Rightarrow$ $p_l^b\leq p_l^a$ is impossible $\Leftrightarrow$ $\mu_l^b\geq \mu_l^a$ is impossible.

\end{itemize}

\subsection{Proof of Theorem 6}
The set of joint feasible strategies $\Pi^p\times \Pi^{pr}\in \mathbb{R}^{N^2}$ is a sublattice of $\mathbb{R}^{N^2}$. The set is also compact since for a power player $n\in \mathcal{N}^p$, $p_n\in\left[P_n^{\min}, P_n^{\max}\right]$ while for a price player $\pi_{m,n}\in\left[\min_{\vec{p}\in\Pi^{p}}\hat{\pi}_{m,n}\left(\vec{p}\right),0\right]$ and the lowest endpoint of the interval is $>-\infty$ for $P_n^{\min}\geq \epsilon>0$, $\forall n$ (see the expression for the success probability (\ref{ITW09:eq03}) and its derivative (\ref{ITW09:partial1}).

Since the set of feasible strategies for power player $n$ is a compact subset of $\mathbb{R}^1$ the payoff function in (\ref{ITW09:eq30}) is supermodular in $p_n$. We have further seen in property \textbf{P.5} that the logarithm of the success probability function of user $n$ has increasing differences for each pair $\left(p_n,p_m\right),\forall m\neq n$ and constant differences for each pair $\left(p_{m_1},p_{m_2}\right)$, $m_1\neq n$, $m_2\neq n$. Then $\log q_n\left(\vec{p},\mu_n\right)$ is supermodular in $\vec{p}\in\Pi^p$. Observe that $p_n\cdot \sum_{m\neq n}\hat{\pi}_{m,n}$ is also supermodular and by property \cite[Lemma 2.6.1(b)]{Topkis} the sum of supermodular functions is supermodular. We reach the conclusion that $J_n = \omega_n\mu_n\cdot \log q_n\left(\vec{p},\mu_n\right)+c_n p_n$ has increasing differences in all pairs $\left(p_i,p_j\right)$ for distinct $i,j\in\mathcal{N}^p$. 

The expression for $J_n$ in (\ref{ITW09:eq30}) is a valuation (has constant differences) for each pair $\left(\hat{\pi}_{i,n},\hat{\pi}_{j,n}\right)$, $i,j\neq n$. Finally for the pairs $\left(p_i,\hat{\pi}_{j,n}\right)$ the function has also increasing differences if $i=n$ and is a valuation for $i\neq n$. Then we reach the conclusion that $J_n$ has increasing differences for each pair $\left(x_i,x_j\right)$, $i,j \in\mathcal{N}^p\cup\mathcal{N}^{pr}$. By definition \ref{ITW09:Def2} the game is supermodular.

The set of feasible joint startegies $\Pi^{p}\times \Pi^{pr}$ is shown to be compact. Observing that the expression in (\ref{ITW09:eq03}) and hence $J_n$ is continuous in $p_n$ and having proven that the game is supermodular, the 'Furthermore' part of the theorem comes from  \cite[Th.4.2.1]{Topkis}.

\subsection{Proof of Theorem 7}
The 'if' part comes directly from the way the extended supermodular game was formulated. For the 'only if' part we argue as follows. Suppose $\vec{p}^e$ is a Nash Equilibrium of the problem. Then for each $n$

\begin{eqnarray}
\label{ITW09:eq35}
p_n^e &= & {\arg\max}_{p_n\in\left[P_n^{\min}, P_n^{\max}\right]} \left\{\omega_n\mu_n\log q_n\left(p_n;\vec{p}_{-n}^e,\mu_n\right)+p_n\sum_{m\neq n}\hat{\pi}_{m,n}^e\right\}\nonumber\\
&= &{\arg\max}_{p_n\in\left[P_n^{\min}, P_n^{\max}\right]}\left\{\omega_n\mu_n \log q_n\left(p_n;\vec{p}_{-n}^e,\mu_n\right)+p_n\sum_{m\neq n} \frac{\omega_m\mu_m}{q_n\left(\vec{p}^e,\mu_n\right)}\frac{\partial q_m\left(\vec{p}^e,\mu_m\right)}{\partial p_n}\right\}\nonumber\\
&= &{\arg\max}_{p_n\in\left[P_n^{\min}, P_n^{\max}\right]}\left\{\omega_n\mu_n \log q_n\left(p_n;\vec{p}_{-n}^e,\mu_n\right)\cdot q_n\left(\vec{p}^e,\mu_n\right)+p_n\sum_{m\neq n} \omega_m\mu_m\frac{\partial q_m\left(\vec{p}^e,\mu_m\right)}{\partial p_n}\right\}\nonumber
\end{eqnarray}
The necessary and sufficient optimality conditions for the above problem are
\begin{eqnarray}
\label{ITW09:eq36}
\omega_n\mu_n \frac{\partial q_n\left(p_n;\vec{p}_{-n}^e,\mu_n\right)}{\partial p_n}\cdot \frac{q_n\left(\vec{p}^e,\mu_n\right)}{q_n\left(p_n,\vec{p}_{-n}^e,\mu_n\right)}+\sum_{m\neq n} \omega_m\mu_m\frac{\partial q_m\left(\vec{p}^e,\mu_m\right)}{\partial p_n} + \nu_n^l-\nu_n^u = 0\nonumber\\
\nu_n^l\cdot \left(P^{\min}_n-p_n\right) \geq 0\ \& \ \nu_n^u\cdot \left(p_n-P^{\max}_n\right) \geq 0\nonumber
\end{eqnarray}
Since $p_n^e$ is the global maximizer (remember that the objective function is concave) the necessary conditions (\ref{ITW09:eq25})-(\ref{ITW09:eq26}) of the scheduling problem are satisfied $\forall n$.

%

\bibliographystyle{unsrt}
\footnotesize
\bibliography{ARQLiterature}

\newpage

\section*{Tables}

\begin{algorithm}
\textbf{Distributed Algorithm for the Scheduling Problem}\\
INITIALIZE
\begin{itemize}
\item Choose the least element of $\Pi^p\times \Pi^{pr}$: $\left(P_n^{\min}\right)$ for the \textit{power} players and $\left(\min_{\vec{p}\in\Pi^p}\hat{\pi}_{m,n}\left(\vec{p}\right)\right)$ for the \textit{price} players.
\item Set $t=0$, $k=0$.
\end{itemize}
REPEAT
\begin{enumerate}
\item \textbf{Power Update:} For $k = 1,\ldots, N$
\begin{itemize}
\item Given $\left(\hat{\pi}_{m,n}^{(t)}\right)$ and $\vec{p}^{(t,k-1)}$
\begin{eqnarray}
p_k^{(t,k)} & = {\arg\max}_{p_k\in\left[P_k^{\min}, P_k^{\max}\right]} J_k^{(t)}\left(p_k;\vec{p}^{(t,k-1)}_{-k}\right)\nonumber
\end{eqnarray}
where $J_k$ is given in (\ref{ITW09:eq30}).
\item $\vec{p}^{(t,k)}_{-k} = \vec{p}^{(t,k-1)}_{-k}$
\end{itemize}

\item \textbf{Price Update:} 
\begin{itemize}
\item For $k = 1,\ldots, N$. Given $\vec{p}^{(t,N)}$ each user $k$ updates the $N-1$ prices $\pi_{k,n}$ for $k\neq n$
\begin{eqnarray}
\pi_{k,n}\left(\vec{p}^{(t,N)}\right) = \omega_k \cdot \mu_k \frac{\partial q_k\left(\vec{p}^{(t,N)},\mu_k\right)}{\partial p_n}\nonumber
\end{eqnarray}
and communicates them to user $n$
\item Each user $n$ receives $N-1$ prices $\pi_{k,n}$ and calculates
\begin{eqnarray}
\hat{\pi}_{k,n}\left(\vec{p}^{(t,N)}\right) = \frac{\pi_{k,n}\left(\vec{p}^{(t,N)}\right)}{q_n\left(\vec{p}^{(t,N)},\mu_k\right)}\nonumber
\end{eqnarray}
\begin{eqnarray}
\label{ITW09:AlgoSum}
c_n^{(t+1)} = \sum_k \hat{\pi}_{k,n}\left(\vec{p}^{(t,N)}\right)
\end{eqnarray}
\end{itemize}

\item Increase $t$ by $1$
\begin{itemize}
\item Set $\vec{p}^{(t,0)} = \vec{p}^{(t-1,N)}$. Set $\left(\hat{\pi}_{m,n}^{(t)}\right) =  \left(\hat{\pi}_{m,n}\left(\vec{p}^{(t-1,N)}\right)\right)$
\end{itemize}

\end{enumerate}
UNTIL $\vec{p}^{(t,0)} = \vec{p}^{(t-1,0)}$ and $\left(\hat{\pi}_{m,n}^{(t)}\right) = \left(\hat{\pi}_{m,n}^{(t-1)}\right)$
\end{algorithm}

\newpage

\section*{Figures}

\newpage

\newpage

\newpage

\end{document}